# Experimental Analysis of Server-Side Caching for Web Performance

Dr. bharat Tripathi [Co-Author]
Dept of Computer Applications.
Allenhouse Business School
Kanpur Uttar Pradesh, India
bharattripathi123@gmail.com

Mohammad Umar [Author]
Dept of Computer Applications
Allenhouse Business School
Kanpur Uttar Pradesh, India
mdumar9140@gmail.com

*Abstract*— **Performance in web applications is a key aspect of user experience and system scalability. Among the different techniques used to improve web application performance, caching has been widely used. While caching has been widely explored in web performance optimization literature, there is a lack of experimental work that explores the effect of simple in-memory caching in small-scale web applications. This paper fills this research gap by experimentally comparing the performance of two server-side web application configurations: one without caching and another with in-memory caching and a fixed time-to-live. The performance evaluation was conducted using a lightweight web server framework, and response times were measured using repeated HTTP requests under identical environmental conditions. The results show a significant reduction in response time for cached requests, and the findings of this paper provide valuable insights into the effectiveness of simple server-side caching in improving web application performance making it suitable for educational environments and small-scale web applications where simplicity and reproducibility are critical**.

*Keywords*— *Server-side caching, Web performance, In-memory cache, Response time optimization, Performance evaluation.*

## I. Introduction

Modern web applications are expected to deliver fast and responsive user experiences while efficiently handling increasing application complexity and user traffic. As a result, server-side performance has become critical for system scalability and user satisfaction. Among various performance optimization techniques, caching is widely used to reduce response time and minimize redundant computations or database queries [1].

In large-scale systems, caching strategies are commonly implemented using distributed caching solutions such as Redis or Memcached [2]. While these approaches are effective in high-traffic environments, small-scale web applications and beginner-level systems often rely on simpler architectures, where deploying and managing distributed caching infrastructure may be unnecessary or complex [3]. This research analyzes the effect of server-side in-memory caching on web application response time.

To achieve this, two functionally identical web applications were implemented: one without caching and another using in-memory caching with a fixed time-to-live (TTL). Both systems were evaluated under identical workloads and environmental conditions. By comparing response times across multiple requests, this study provides a clear and reproducible performance comparison, offering practical insights into the effectiveness and limitations of lightweight server-side caching for web performance optimization.

The primary contribution of this work is a controlled and reproducible experimental evaluation of simple server-side in-memory caching under identical conditions. Unlike prior studies that focuses on distributed or enterprise scale caching systems, this works emphasizes simplicity and practical applicability for small scale applications and educational use cases.

## II. Related Work

Previous studies on web performance optimization emphasize techniques such as HTTP caching, database query optimization, content delivery networks, and distributed caching systems. Research on caching has predominantly focused on enterprise-scale systems, microservices architectures, and cloud-based infrastructures [4],[5] While these comprehensive caching solutions offer significant benefits for large-scale deployments [6], their complexity often overshadows the needs of simpler web applications. For instance, solutions like Redis and Memcached, while effective for distributed environments [5], can introduce overhead for smaller applications where in-memory caching might suffice [1].

Several studies demonstrate that caching significantly reduces latency and improves throughput in high-traffic environments [7]. However, these works often assume the use of external caching systems or complex architectures. Limited attention has been given to evaluating basic in-memory caching techniques in small-scale server environments, particularly for educational or early-stage development scenarios. This research addresses this gap by analyzing server-side in-memory caching in Express.js, focusing on simplicity, reproducibility, and practical applicability.

Despite extensive research on advanced caching mechanisms, there is limited experimental evaluation of light weight in-memory caching under controlled conditions. This study addresses this gap by focusing on a minimal and reproducible experimental setup.

### III. Problem Statement

Although caching is a well-established optimization technique [8], however, the quantitative impact on response time of simple in-memory caching in web applications is often based on irreproducible and biased developer tests, not robust controlled experiments [1]. Additionally, several studies demonstrate the measurable impact of in-memory caching on application latency and response times through controlled experiments [9]. As a result, developers and students often lack clear evidence regarding the benefits and limitations of lightweight server-side caching approaches in such environments.

### IV. Methodology

**A. Experimental setup**

Two web applications were implemented using Node.js with the Express.js framework. Both servers exposed an identical HTTP GET endpoint to ensure functional equivalence. A simulated heavy computation was introduced with a delay to represent database access or processing overhead.

**Server A (Without Caching):**
Each incoming request executed the simulated computation before returning a response.

**Server B (With Caching):**
The server stored the computed response in an in-memory cache with a fixed time-to-live (TTL). Subsequent requests received within the TTL window were served directly from the cache.

All experiments were conducted on the same machine to ensure identical hardware, operating system, and runtime conditions.

**B. Caching Mechanism:** The caching mechanism employs a simple in-memory data structure stored in server RAM. An expiration policy was implemented using a TTL value to invalidate cached data automatically.

**Cache Hit**: The requested response is retrieved directly from the cache [10].

**Cache Miss**: The request is processed normally, and the resulting response is stored in the cache [11].

This approach ensures minimal implementation complexity while enabling a clear comparison between cached and non-cached request handling.

**C. Data Collection**

Response times were measured in milliseconds using timestamps recorded at the beginning and end of request handling. Requests were sent with Postman under identical conditions for both servers.

Each experiment consisted of ten sequential HTTP GET requests sent at fixed intervals using Postman. The same request payload and headers were used for both cached and non-cached servers to ensure consistency. Response time was recorded in milliseconds using server-side timestamps to eliminate client-side measurement bias. The TTL value for the cache was fixed throughout the experiment.

To ensure the reproducibility of the experimental results, the complete server-side source code is publicly available at: **https://github.com/developer-umar/Caching_experiment.**

### V. Experimental Results

**Response Time Analysis:** The server without caching consistently exhibited higher response times due to repeated execution of the simulated heavy computation. In contrast, the cached server demonstrated a significant reduction in response time after the initial request.

Table I: Response Time Comparison Between Cached and Non-Cached Servers

| Request No | No Cache (ms) | Cache (ms) |
|---|---|---|
| 1 | 1012 | 1011 |
| 2 | 1013 | 0 |
| 3 | 1008 | 0 |
| 4 | 1011 | 1009 |
| 5 | 1010 | 0 |
| 6 | 1012 | 0 |
| 7 | 1008 | 0 |
| 8 | 1008 | 0 |
| 9 | 1011 | 0 |
| 10 | 1008 | 1012 |

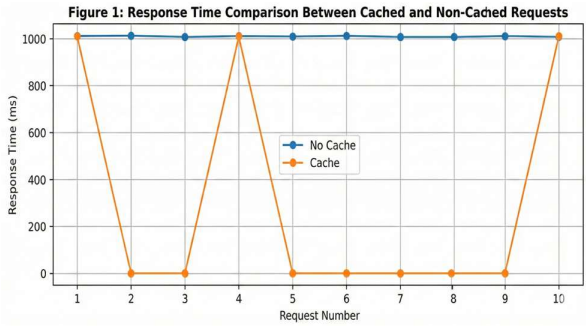

Figure 1: Response Time Comparison Between Cached and Non-Cached Requests

As shown in Table I and Figure 1, **the server without caching consistently exhibits higher response times due to repeated execution of the simulated computation. In contrast, the cached server demonstrates a significant reduction in response time after the initial request**, with subsequent requests served directly from memory.

**Cache Hit and Miss Analysis:** During the experiment, the number of cache hits increased significantly after the first request within the TTL window. This behavior confirms the effectiveness of in-memory caching to reduce redundant computations.

### Table II: Cache Hit and Miss Statistics

| Metric | Count |
| --- | --- |
| Total Requests | 10 |
| Cache Hits | 7 |
| Cache Misses | 3 |

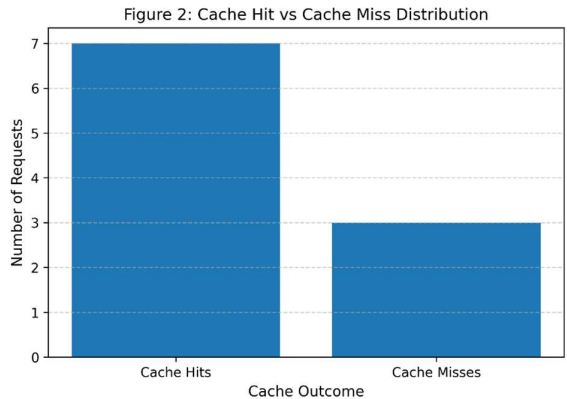

Figure 2: Cache Hit vs Cache Miss Distribution

The cache hit and miss statistics are presented in Table II and Figure 2. **The results indicate that after the initial cache miss, the majority of subsequent requests are served from memory.** This behavior confirms the effectiveness of in-memory caching to reduce redundant computations and improve response time.

On average, the cached responses **showed a 90-95% reduction in response time** compared to the non-cached responses after the initial request.

### VI. Discussion

The results indicate that even simple in-memory caching can substantially reduce response time in small-scale web applications [12]. Cached requests were served significantly faster than non-cached requests, as redundant computation was avoided after the initial cache population [13]. This demonstrates that lightweight caching can be an effective optimization strategy.

Despite these benefits, the proposed approach has limitations. In-memory caching is volatile, and cached data is lost upon server restart [14]. Furthermore, this technique is not suitable for distributed or multi-instance systems, where cache consistency across nodes is required [15]. Applications involving real-time or highly dynamic data may also experience consistency challenges when using simple in-memory caching [16].

The results suggest that while lightweight server-side caching is highly effective at improving response time in small-scale and single-instance applications, it should be carefully evaluated before applying to larger or distributed system architectures. Compared to distributed caching systems discussed in prior work, the proposed approach lower implementation overhead while still achieving significant performance gains in controlled environments.

### VII. Limitations

• The caching system is based on memory storage, and the cache is not persistent across server restarts [17].

• The experimental study is conducted in the context of a single-instance server setup and does not consider distributed server scenarios.

• The possible implications of data consistency issues in real-time or dynamic content are not considered.

• The workload used in the experimental studies is simulated and may not accurately reflect real-world traffic patterns.

### VIII. Conclusion

This study evaluated the impact of in-memory caching on response time in web applications with a controlled and reproducible setup. The results confirm that lightweight caching significantly improves performance for repeated requests by reducing redundant computation. While this approach is not suitable for large-scale or distributed systems, it provides an effective, easy-to-implement optimization strategy for small-scale web applications and educational use cases.

## IX. Future Work

Future studies can be conducted to further this research by exploring the integration of distributed caching systems, such as Redis, to determine scalability in multi-instance server environments. Furthermore, system performance can be tested in concurrent and high-load request scenarios to more accurately represent real-world usage patterns. Comparative analysis of various caching approaches and cache eviction strategies can also further illustrate the effect of caching on response time and resource usage. Finally, future studies can investigate memory usage, cache efficiency, and overall system scalability to gain a more complete understanding of server-side caching performance.